\documentclass[letterpaper]{jpconf}
\usepackage{graphicx}
\begin{document}

\begin{flushright}
{\bf UCHEP-04-08}
\end{flushright}

\title{Properties of the {\boldmath $D_{sJ}$} states}

\author{Alexey Drutskoy \\ {\it \small (on behalf of the Belle Collaboration)}}

\address{Physics Department, University of Cincinnati, 345 College Court,
Cincinnati, OH 45221, USA}

\ead{drutskoy@physics.uc.edu}

\begin{abstract}
Recent measurements involving the newly discovered $D_{sJ}$ particles 
are reported. The results of $D_{sJ}$ production and decay branching 
fraction measurements are shown. Possible spin-parity and quark content 
assignments of $D_{sJ}$ mesons are discussed. The results are based 
on a large data sample recorded by the Belle detector at 
the KEKB $e^+ e^-$ collider.
\end{abstract}.

\section{Introduction}

A narrow $D_{sJ}^\ast(2317)^+$ resonance decaying to the $D_s \pi^0$
final state has been observed by the BaBar collaboration \cite{baba}
in $e^+ e^-$ continuum interactions. Later the CLEO collaboration 
observed a nearby narrow $D_{sJ}(2460)^+$ resonance
decaying to the $D^*_s \pi^0$ final state \cite{clea}. The Belle
experiment has confirmed the existence of 
these two resonances \cite{bela,belb}.
Comparing the measured $D_{sJ}$ decay branching fractions and upper limits,
the quantum numbers have been classified tentatively
as $J^P=0^+$ for $D_{sJ}^\ast(2317)^+$ and $J^P=1^+$ for $D_{sJ}(2460)^+$.
However, the measured masses of these resonances are significantly lower
than the values predicted within potential models for $0^+$ 
and $1^+$ states~\cite{dsjaa}.
There has been a significant effort to explain the surprising
$D_{sJ}$ masses~\cite{dsjaa},
and some authors have discussed the possibility of
four-quark content in the $D_{sJ}$ states.
To clarify the behaviors of the $D_{sJ}$ states the Belle collaboration 
has searched for $B$ decays to two-body final states involving 
the combination of a $D_{sJ}^+$ meson and a $D$, $D^*$, kaon or pion.

\section{{\boldmath $B \to \bar{D}^{(*)} D_{sJ}$} decays}

With a data sample of 253~fb$^{-1}$ collected with the Belle
detector, the $B \to \bar{D} D_{sJ}$ decay modes are measured with 
improved accuracy (Belle previously published the observation 
of these decays with 140~fb$^{-1}$ of data~\cite{belb})
and new decay modes $B \to \bar{D}^* D_{sJ}$ are observed.
These processes are described by conventional tree diagrams,
similar to the $B \to \bar{D}^{(*)} D_s$ decays.
$D_{sJ}$ candidates are reconstructed 
in the modes $D_s^{(*)} \pi^0$, $D_s^{(*)} \gamma$ and $D_s^{(*)} \pi^+\pi^-$.

The results of  
the combined data fit (isospin invariance is assumed to combine decay channels) are listed in
Table 1. From these measurements we obtain the branching fraction ratio:
${\cal B}(D_{sJ}(2460)^+ \to D_s^+ \gamma)/{\cal B}(D_{sJ}(2460)^+ \to D_s^{*+}\pi^0) = 0.43 \pm 0.08 \pm 0.04$.

The efficiency-corrected helicity angle\,\cite{belb} distributions 
are consistent with expectations 
for the $J=0$ hypothesis for $D_{sJ}^\ast(2317) \to D_s \pi^0$ decay (Fig.~1, left)
and with expectations for the $J=1$ hypothesis for
$D_{sJ}(2460) \to D_s \gamma$ decay \mbox{(Fig.~1, right)}.

\begin{center}
\begin{table}[t]
\caption{Branching fractions and signal significances for the combined fit results.}
\centering
\begin{tabular}{@{}lcc@{}}
\br
Decay channel & ${\cal B}$, 10$^{-4}$ & Significance \\
\mr
$B \to \bar{D} D_{sJ}^\ast(2317)\ [D_s \pi^0]$ & $10.1 \pm 1.5 \pm 3.0$ & 9.5$\sigma$ \\
$B \to \bar{D} D_{sJ}^\ast(2317)\ [D_s^* \gamma]$ & $4.0^{+1.5}_{-1.4} (<8.4)$ & 3.5$\sigma$ \\
$B \to \bar{D} D_{sJ}(2460)\ [D_s^* \pi^0]$ & $14.8^{+2.8}_{-2.5} \pm 4.4$ & 8.6$\sigma$ \\
$B \to \bar{D} D_{sJ}(2460)\ [D_s \gamma]$ & $6.4 \pm 0.8 \pm 1.9$ & 11$\sigma$ \\
$B \to \bar{D} D_{sJ}(2460)\ [D_s^* \gamma]$ & $2.6^{+1.1}_{-1.0} (<5.7)$ & 3.0$\sigma$ \\
$B \to \bar{D} D_{sJ}(2460)\ [D_s \pi^+\pi^-]$ & $1.0^{+0.5}_{-0.4} (<2.3)$ & 2.6$\sigma$ \\
$B \to \bar{D} D_{sJ}(2460)\ [D_s \pi^0]$ & $0.2^{+0.7}_{-0.5} (<1.7)$ & - \\
\mr
$B \to \bar{D}^* D_{sJ}^\ast(2317)\ [D_s \pi^0]$ & $3.1^{+2.1}_{-1.7} (<8.5)$ & 2.0$\sigma$ \\
$B \to \bar{D}^* D_{sJ}(2460)\ [D_s^* \pi^0]$ & $28.7^{+7.4}_{-6.4} \pm 8.6$ & 6.9$\sigma$ \\
$B \to \bar{D}^* D_{sJ}(2460)\ [D_s \gamma]$ & $12.7^{+2.2}_{-2.0} \pm 3.8$ & 10$\sigma$ \\
\br
\end{tabular}
\end{table}
\end{center}

\begin{figure}[h]
\begin{minipage}{36pc}
\begin{center}
\includegraphics[width=5.3cm]{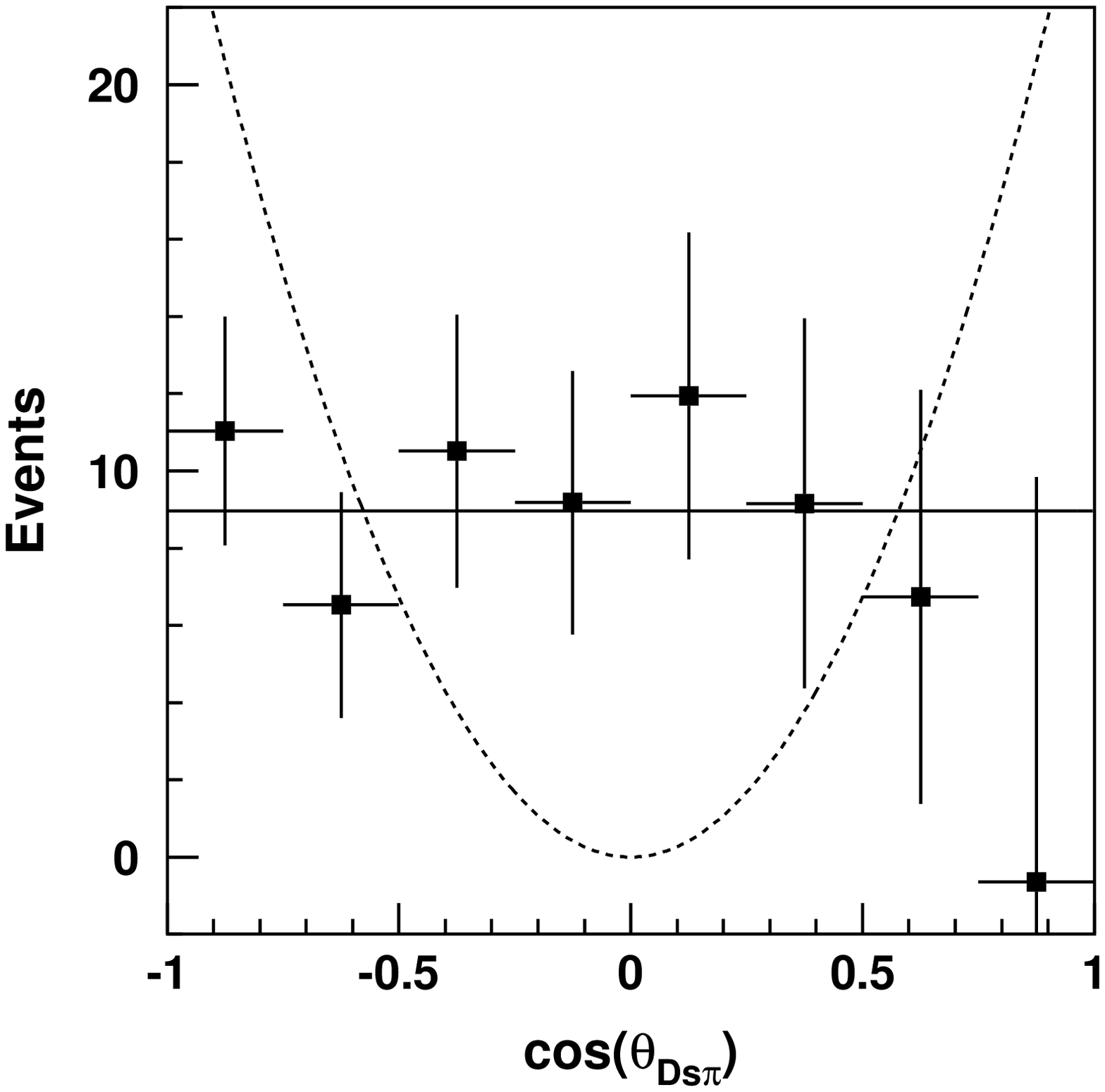}\includegraphics[width=5.3cm]{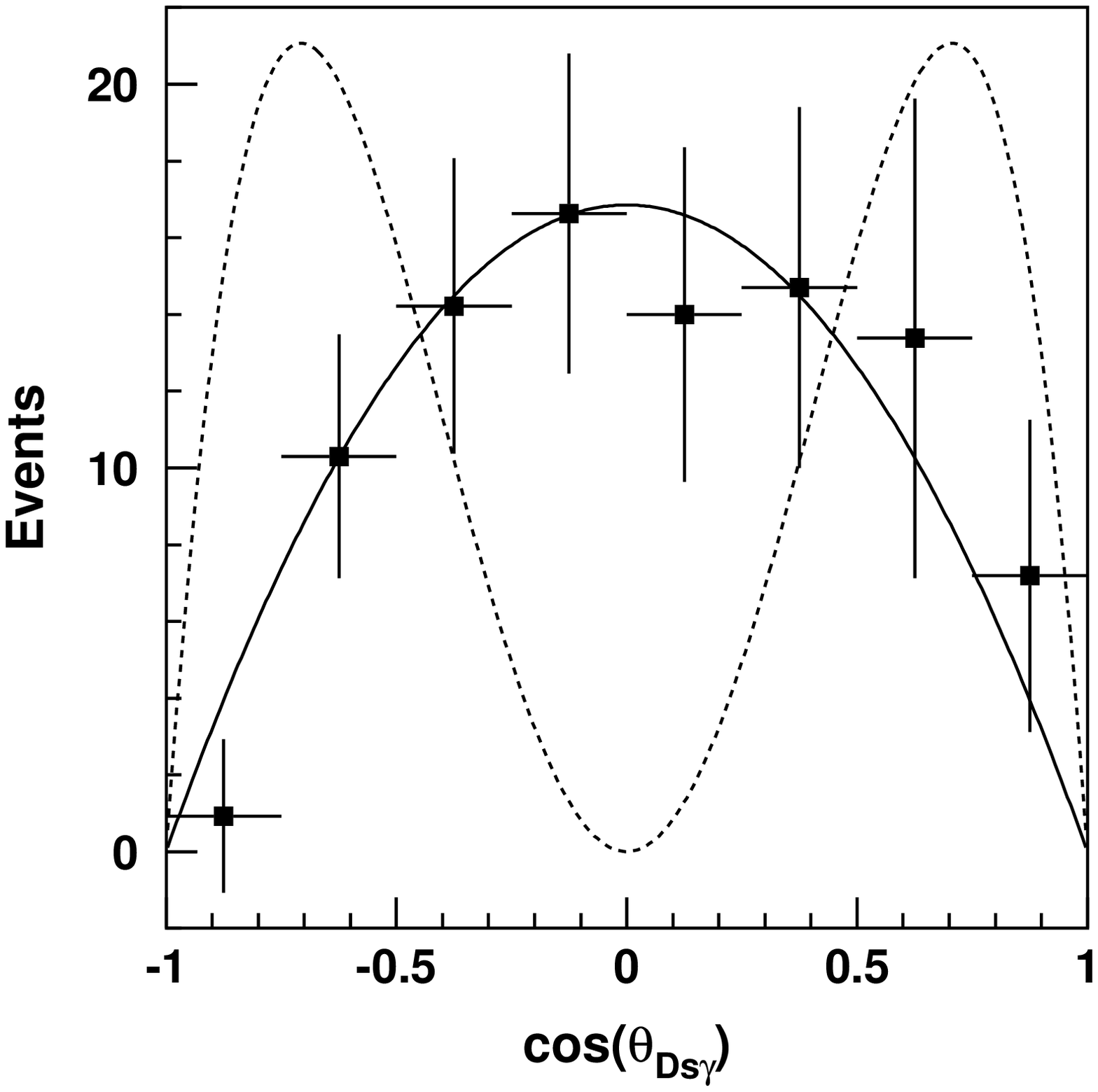}
\end{center}
\caption{
The efficiency-corrected cos\,$\theta_{\rm\,hel}$ distributions for 
$D_{sJ}(2317) \to D_s \pi^0$ (left) 
and $D_{sJ}(2460) \to D_s \gamma$ (right) decays. Solid and dashed curves are 
predictions: for the $J=0$ and $J=1$ hypotheses of $D_{sJ}^\ast(2317)$; $J=1$ and $J=2$
hypotheses of $D_{sJ}(2460)$, respectively.}
\end{minipage} 
\end{figure}

\section{{\boldmath $\bar{B}^0 \to D_{sJ}^+ K^-$} and {\boldmath $\bar{B}^0 \to D_{sJ}^- \pi^+$} decays}
\smallskip

The decays \mbox{$\bar{B}^0 \rightarrow D_{sJ}^+ K^-$} and 
\mbox{$\bar{B}^0 \rightarrow D_{sJ}^- \pi^+$} are studied by the
Belle collaboration for the first time with 140~fb$^{-1}$ of data.
The $\Delta M(D_{sJ}) \equiv M(D_s^+ \pi^0(/\gamma)) - M(D_s^+)$ distributions
for the various $D_{sJ}^+ K^-$ and $D_{sJ}^- \pi^+$ combinations
are shown in Fig.~2 for candidates from the $B$ signal region.
A clear $\bar{B}^0 \to D_{sJ}^\ast(2317)^+ K^-$ 
signal is observed; no significant signals are observed in the remaining modes.
The branching fractions, upper limits and significances for these decay 
channels are listed in Table~2.

The \mbox{$\bar{B}^0 \rightarrow D_{sJ}^+ K^-$} decay
is generally described by the $W$ exchange diagram (Fig.~3a) or, 
alternatively, the tree diagram with final state interaction (Fig.~3b).
If the $D_{sJ}$ mesons have a four-quark component,
then the tree diagram with $s\bar{s}$ pair 
creation (shown in Fig.~3c) may also contribute.

Assuming the approximate values of $D_{sJ}$ decay branching fractions
($\mathcal{B}(D_{sJ}^\ast(2317)^+ \to D_s^+ \pi^0) \sim 100\%$,
$\mathcal{B}(D_{sJ}(2460)^+ \to D_s^+ \gamma) \sim 30\%$),
we conclude that $\mathcal{B}(\bar{B}^0 \to D_{sJ}^\ast(2317)^+ K^-)$ 
is of the same order of magnitude as $\mathcal{B}(\bar{B}^0 \to D_s^+ K^-)$
but at least a factor of two larger than
$\mathcal{B}(\bar{B}^0 \to D_{sJ}(2460)^+ K^-)$.

\begin{figure}[h]
\begin{minipage}{36pc}
\begin{center}
\includegraphics[width=6cm]{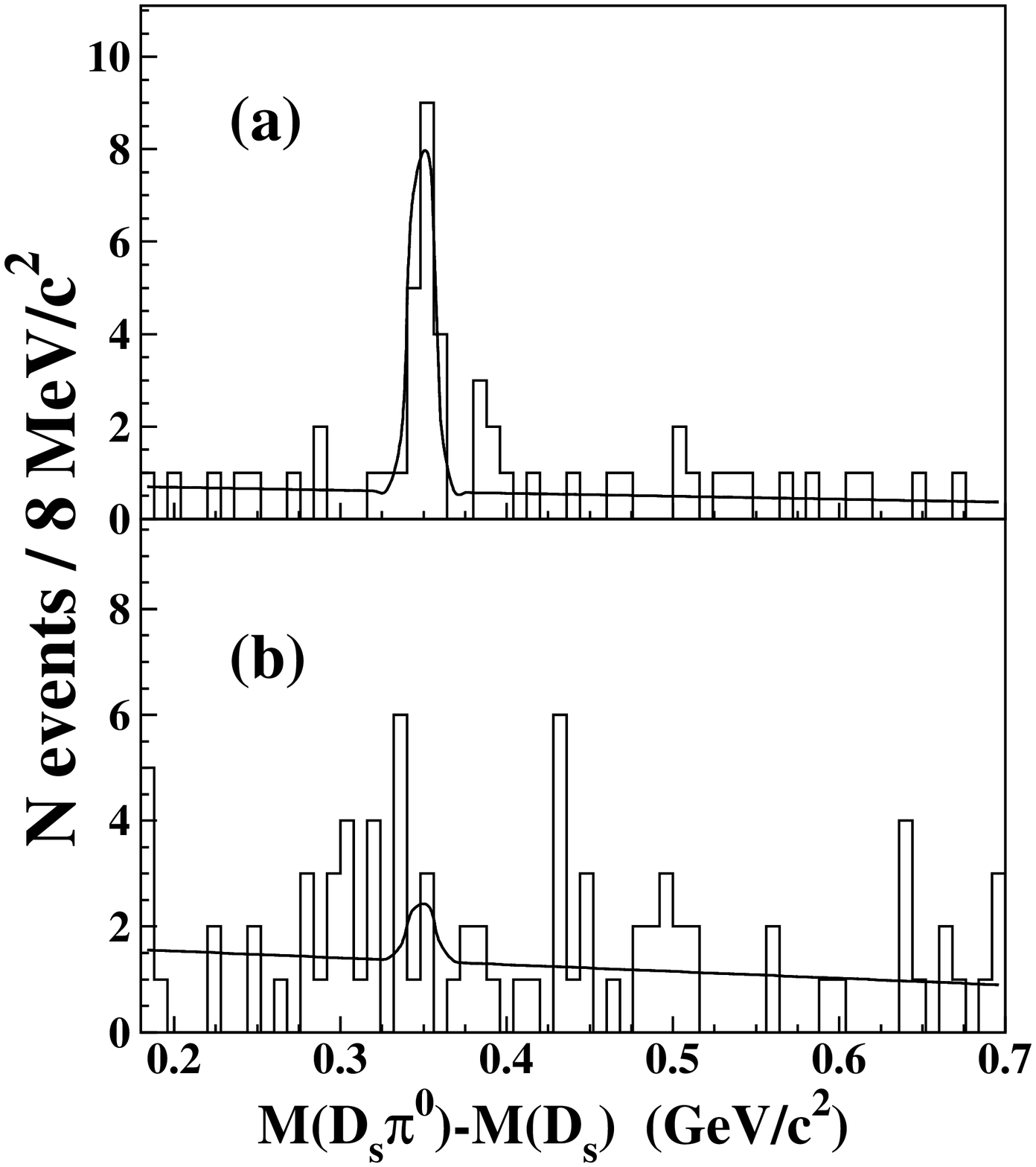}\includegraphics[width=6cm]{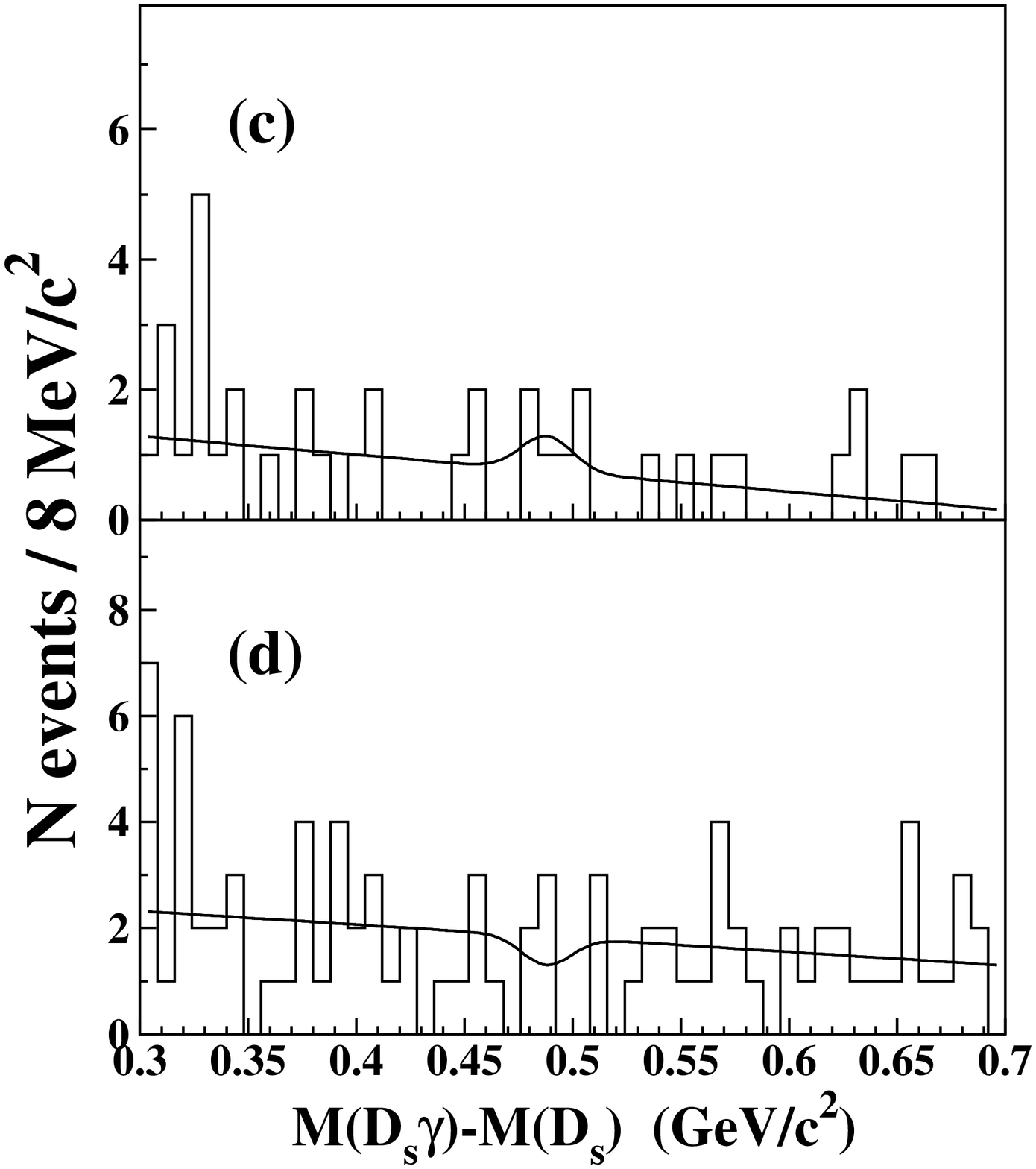}
\end{center}
\caption{
$\Delta M(D_{sJ})$ distributions for $\bar{B}^0$ decays to 
(a) $D_{sJ}^\ast(2317)^+ K^-$, 
\newline (b) $D_{sJ}^\ast(2317)^- \pi^+$, (c) $D_{sJ}(2460)^+ K^-$ 
and (d) $D_{sJ}(2460)^- \pi^+$.}
\end{minipage} 
\end{figure}

\vspace{-0.3mm}
\begin{figure}[h]
\begin{minipage}{36pc}
\begin{center}
\includegraphics[width=5.3cm]{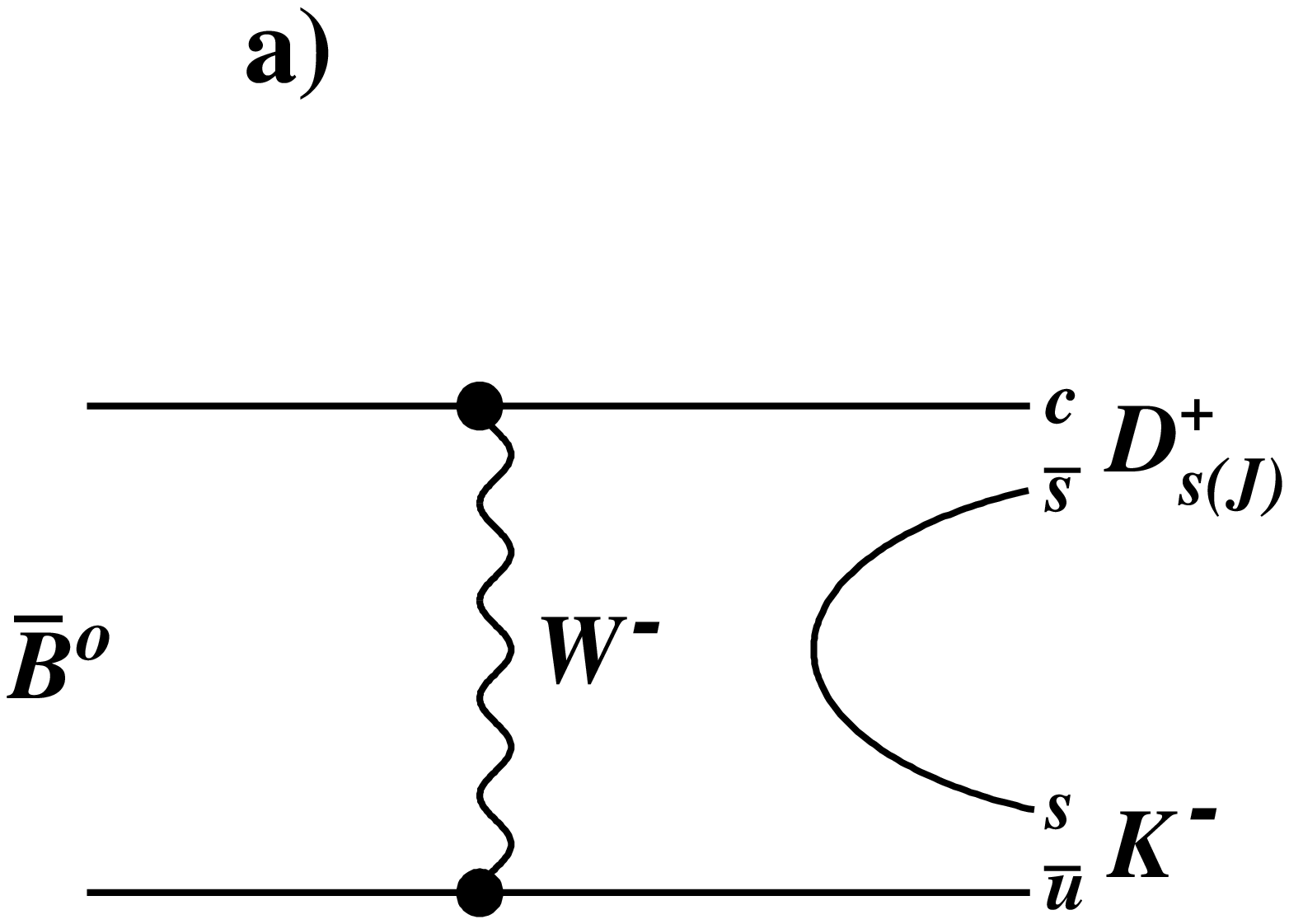}\includegraphics[width=5.3cm]{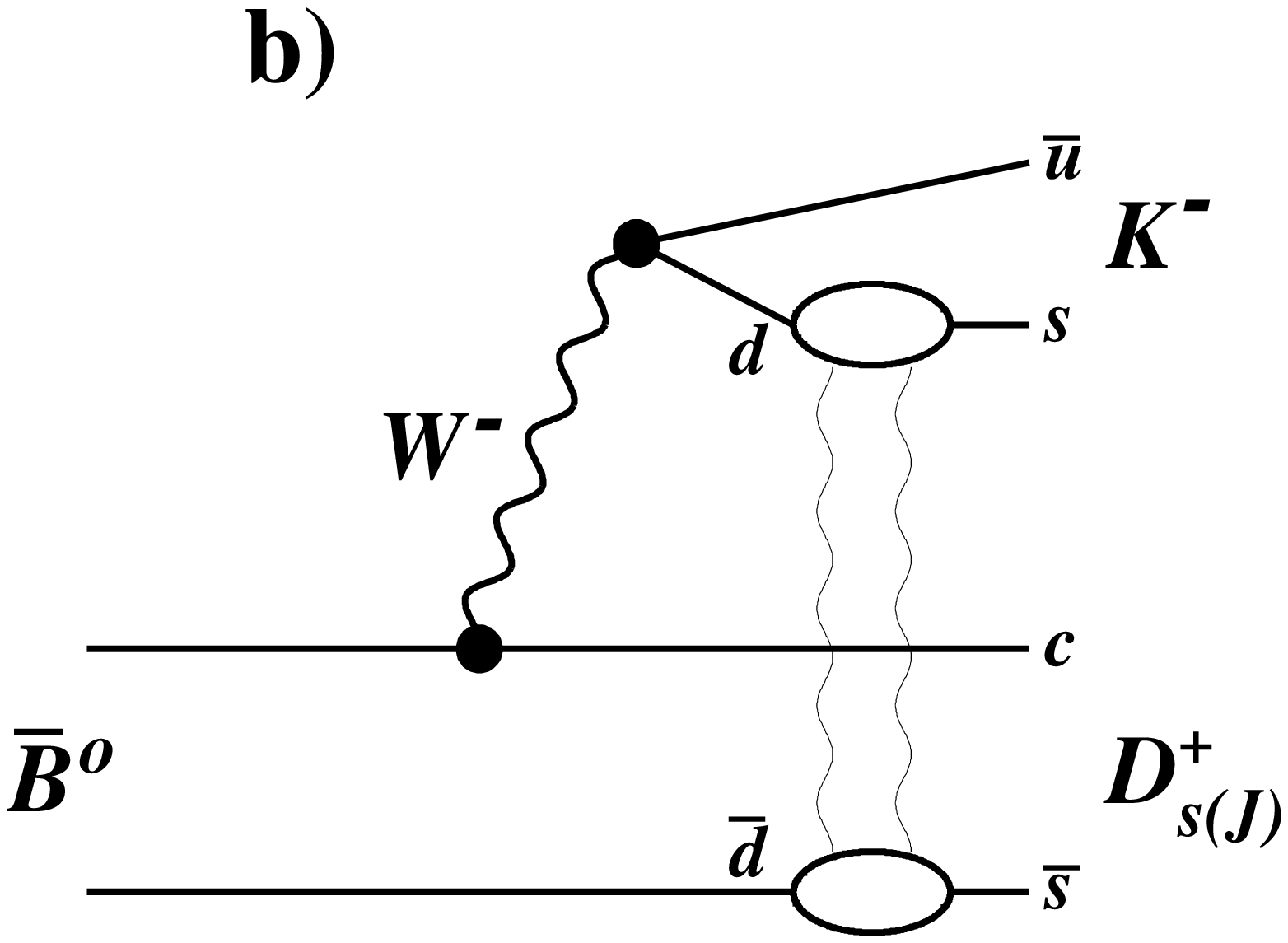}\includegraphics[width=5.3cm]{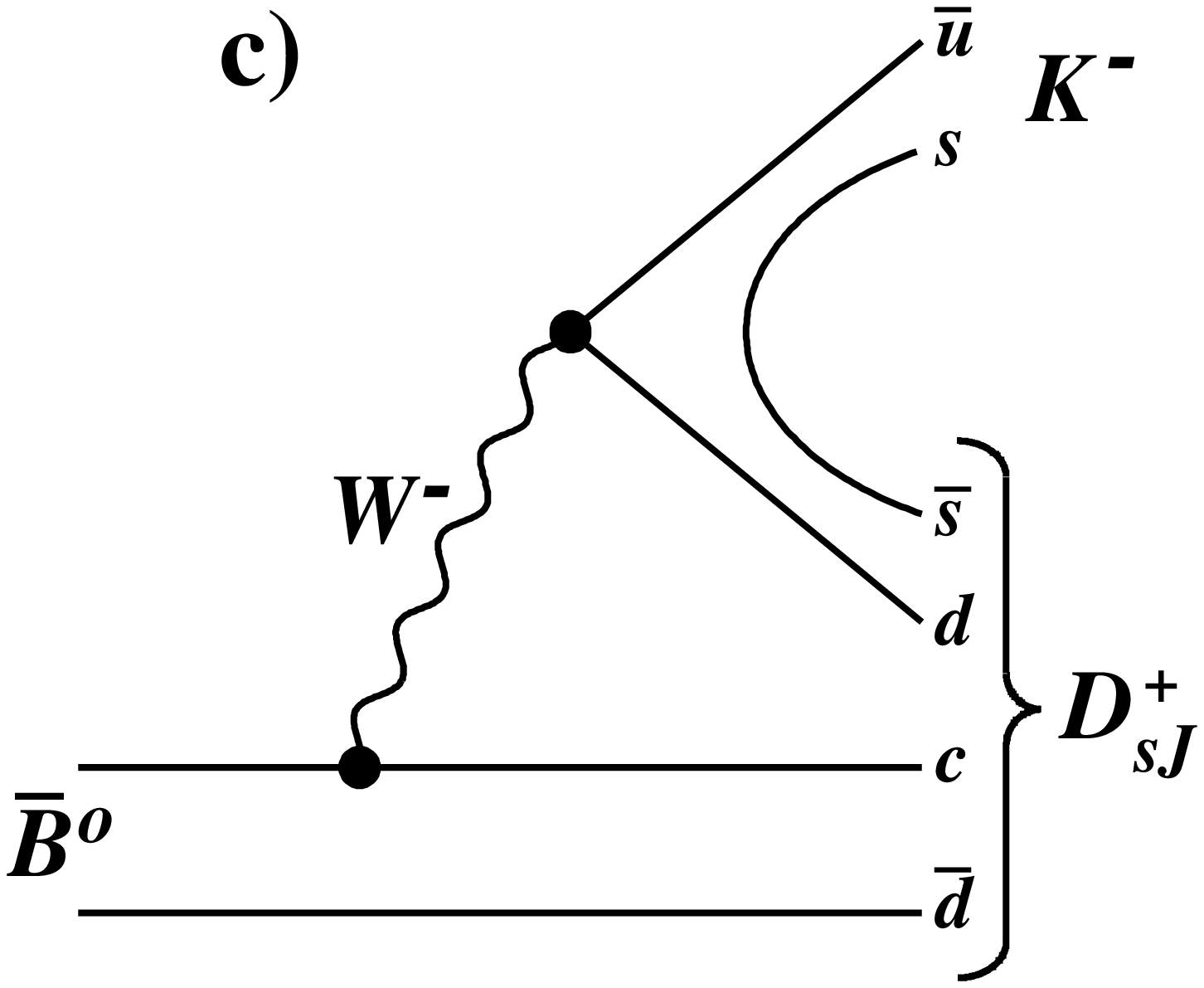}
\end{center}
\caption{ Diagrams describing $\bar{B}^0 \to D_{sJ}^+ K^-$ decay.}
\end{minipage} 
\end{figure}

\begin{center}
\begin{table}[h]
\caption{Branching fractions and signal significances for the $B$ decays to
$D_{sJ}^+ K^-$ and $D_{sJ}^- \pi^+$ final states.}
\centering
\begin{tabular}{@{}lcc@{}}
\br
Decay channel & ${\cal B}$, 10$^{-5}$ & Significance \\
\mr
$\bar{B}^0 \to D_{sJ}^\ast(2317)^+ K^-\ \ [D_s \pi^0]$ & $5.3^{+1.5}_{-1.3} \pm 0.7 \pm 1.4$ & 6.8$\sigma$ \\
$\bar{B}^0 \to D_{sJ}^\ast(2317)^- \pi^+\ \ [D_s \pi^0]$ & $<2.5\ (90\%$ C.L.) & - \\
$\bar{B}^0 \to D_{sJ}(2460)^+ K^-\ \ [D_s \gamma]$ & $<0.94\ (90\%$ C.L.) & - \\
$\bar{B}^0 \to D_{sJ}(2460)^- \pi^+\ \ [D_s \gamma]$ & $<0.40\ (90\%$ C.L.) & - \\
\br
\end{tabular}
\end{table}
\end{center}

\section{Quantum numbers of {\boldmath $D_{sJ}$} resonances}
\smallskip

To determine the $D_{sJ}$ quantum numbers, the following
experimental results are considered:

\begin{itemize}
\item The decay mode $D_{sJ}^\ast(2317)^+ \to D_s^+ \pi^0$
is dominant. No significant signals are observed for the
$D_{sJ}^\ast(2317)^+ \to D_s^+ \gamma$ and
$D_{sJ}^\ast(2317)^+ \to D_s^{*+} \pi^0$ decays.
\item Significant signals are observed in the
decay modes $D_{sJ}(2460)^+ \to D_s^+ \gamma$
and $D_{sJ}(2460)^+ \to D_s^{*+} \pi^0$.
No significant signal is observed for the
$D_{sJ}(2460)^+ \to D_s^+ \pi^0$ decay.
\item The helicity angular distributions in the
$B \to \bar{D}^{(*)} D_{sJ}$ decays favor the $J = 0$
hypothesis for $D_{sJ}^\ast(2317)^+$ and $J = 1$
for $D_{sJ}(2460)^+$.
\item The helicity angular distribution in the
$\bar{B}^0 \to D_{sJ}^+ K^-$ decay favors the $J = 0$
hypothesis for $D_{sJ}^\ast(2317)^+$.
\end{itemize}

Taking into account this information, the $0^+$
quantum numbers for the $D_{sJ}^\ast(2317)^+$ and $1^+$
for the $D_{sJ}(2460)^+$ are now established
with high confidence.

\section{Quark content of {\boldmath $D_{sJ}$} resonances}
\smallskip

In recent theoretical papers many authors indicate that 
there are no substantial reasons to assume a four-quark content
(or significant admixture) for the $D_{sJ}$  mesons. 
However, there are some experimental results, which
can not be explained clearly within a two-quark picture:

\smallskip
{\bf 1)} The branching fraction ratio measured by the Belle
collaboration in continuum:
${\cal B}(D_{sJ}^\ast(2317)^+ \to D_s^{*+} \gamma)/{\cal B}((D_{sJ}^\ast(2317)^+ \to D_s^+\pi^0) < 0.18\ (at\ 90\%\ C.L.)$.
\smallskip

{\bf 2)} The branching fraction ratio 
${\cal B}(D_{sJ}(2460)^+ \to D_s^+ \pi^+ \pi^-)/{\cal B}((D_{sJ}(2460)^+ \to D_s^{*+} \pi^0)$ 
is measured by the Belle collaboration to be $0.14 \pm 0.04 \pm 0.02$
in continuum and \mbox{$< 0.13\ (at\ 90\%\ C.L.)$} in $B$ meson decays.
\smallskip

{\bf 3)} The branching fractions for $B \to \bar{D}^{(*)} D_{sJ}$
measured by Belle are an order of
magnitude smaller than those for $B \to \bar{D}^{(*)} D_s$.
\smallskip

{\bf 4)} The branching fraction for $\bar{B}^0 \to D_{sJ}^\ast(2317)^+ K^-$
measured by Belle is of the same order of magnitude as that for 
$\bar{B}^0 \to D_s^+ K^-$.
\smallskip

In case of a conventional two-quark interpretation,
the $D_{sJ}$ decays with $\pi^0$ in the final state must be suppressed
due to isospin violation, and the 
ratios 1) and 2) are expected to be somewhat larger
than the obtained values \cite{haya}.
However, these values are readily explained within 
a four-quark $D_{sJ}$ interpretation \cite{haya}.
The order-of-magnitude difference in the $B \to \bar{D}^{(*)} D_{sJ}$
and $B \to \bar{D}^{(*)} D_s$ decay branching fractions
indicated in 3) can be also explained as the effect of an additional 
quark pair creation in a four-quark interpretation \cite{chen}.
(It has to be mentioned that, due to vector current conservation,
$B$ decay tree diagrams 
with a $0^+$ meson produced from the virtual $W$ boson are expected to be
suppressed \cite{supr}; this problem was not discussed in \cite{chen}.)
In contrast to 3), the branching fractions indicated in 4) are
of the same order, but it may be 
explained by the contribution from the diagram shown in Fig.~3c.
Further theoretical analysis of the
two- and four-quark interpretation opportunities will be required
to understand the nature of the $D_{sJ}$ mesons.

\smallskip

\end{document}